# EllipsoNet: Deep-learning-enabled optical ellipsometry for complex thin films

*Ziyang Wang[†], Yuxuan Cosmi Lin[‡, 1, \*], Kunyan Zhang[₴], Wenjing Wu[†,%], Shengxi Huang[†,\*]*

[†] Department of Electrical and Computer Engineering, Rice University, Houston, TX 77005, USA

[‡] Department of Electrical Engineering and Computer Sciences, University of California, Berkeley, CA 94720, USA

[₴] Department of Electrical Engineering, The Pennsylvania State University, University Park, PA 16802, USA

[%] Applied Physics Graduate Program, Smalley-Curl Institute, Rice University, Houston, TX 77005, USA

[1] Z.W. and Y.C.L. should be considered joint first author

**ABSTRACT**

Optical spectroscopy is indispensable for research and development in nanoscience and nanotechnology, microelectronics, energy, and advanced manufacturing. Advanced optical spectroscopy tools often require both specifically designed high-end instrumentation and intricate data analysis techniques. Beyond the common analytical tools, deep learning methods are well suited for interpreting high-dimensional and complicated spectroscopy data. They offer great opportunities to extract subtle and deep information about optical properties of materials with simpler optical setups, which would otherwise require sophisticated instrumentation. In this work, we propose a computational ellipsometry approach based on a conventional tabletop optical microscope and a deep learning model called *EllipsoNet*. Without any prior knowledge about the multilayer substrates, EllipsoNet can predict the complex refractive indices of thin films on top of these nontrivial substrates from experimentally measured optical reflectance spectra with high accuracies. This task was not feasible previously with traditional reflectometry or ellipsometry methods. Fundamental physical principles, such as the Kramers-Kronig relations, are spontaneously learned by the model without any further training. This approach enables in-operando optical characterization of functional materials within complex photonic structures or optoelectronic devices.

**INTRODUCTION**

Complex refractive indices are among the most fundamental properties of materials. They act as optical "fingerprints" of materials and contain rich information about light-matter interaction, such as optical interband/intraband transitions, quantum confined state transitions, phonon polaritons, and exciton polaritons. They are also essential material properties in designing photonic and optoelectronic devices.[1–6] Ellipsometry is a widely used technique to measure the refractive index of thin films.[7–9] It first measures the changes in polarization in terms of the amplitude ratio Ψ and phase difference Δ (**Figure S1**). The measured Ψ and Δ are related to the optical reflectance ratio between p and s polorizations, $\frac{r_p}{r_s} = \tan(\Psi)\, e^{i\Delta}$, which is a function of the thickness and complex refractive index. Ψ and Δ will be fitted by a model that describes the multilayer sample, where the refractive index of the target layer consists of multiple oscillators. Therefore, the refractive indices of the target layer can be obtained (**Figure S2**). Despite its wide use, ellipsometry technique still faces three main challenges. First, selecting a proper model requires intervention by human experts due to intensive parameter tuning, including the selection of type and number of oscillators.[10] Further, the substrate structures have to be simple and known. However, in many practical scenarios, the substrate structures are inevitably more complicated with partially unknown information. Finally, the optical setup for ellipsometry demands a large incident angle which requires special instrumentation and is not directly implementable on a common reflection optical microscope setup. Previous studies have made efforts to simplify the process of parameter selections but were still limited by the requirements of simple and well-defined substrate structures, and unconventional optical setups.[11–13]

In this work, we developed a deep learning method that extracts complex refractive indices of thin films placed on unknown and arbitrary multilayer substrates from optical reflectance spectra measured on an optical microscope. Unlike traditional reflectometry or ellipsometry, our approach does not require extensive fitting and is able to tackle all the three challenges mentioned above. First, our framework obtains refractive indices of thin films without solving inverse functions or tuning parameters. Second, our method can be used in complex substrate structures without knowing the structure parameters or materials of substrates. Moreover, our model only takes reflectance spectra as the inputs, which can be easily integrated with optical microscopes (**Figure 1a**). Specifically, we designed an encoder-decoder convolutional neural network named EllipsoNet that takes reflectance spectra as the inputs and predicts the corresponding refractive indices (**Figure 1a**, **b**, **c**). To train the neural network, we generated a dataset using over 400 materials with density functional theory (DFT) simulated refreactive indeices and 450,000 multilayer stack structures. With an independently generated dataset of testing materials and multilayer stack structures, the predictions made by EllipsoNet reach an overall median Pearson's correlation coefficient of 0.88. We further validated our method using experimentally measured reflectance of real 2D materials on different substates. We also showed that a more complex version of EllipsoNet, called C-EllipsoNet, can deal with even more complicated substrate structures. Finally, we found that both EllipsoNet and C-EllipsoNet spontaneously learn the Kramers-Kronig (KK) relations, a fundamental physical principle governing the light-matter interaction, without any extra training.[14] EllipsoNet will enable in-operando characterization of unknown materials in complex photonic structures. Our deep neural network approach can also be extended to extract other material properties and be applied to various spectroscopic data.

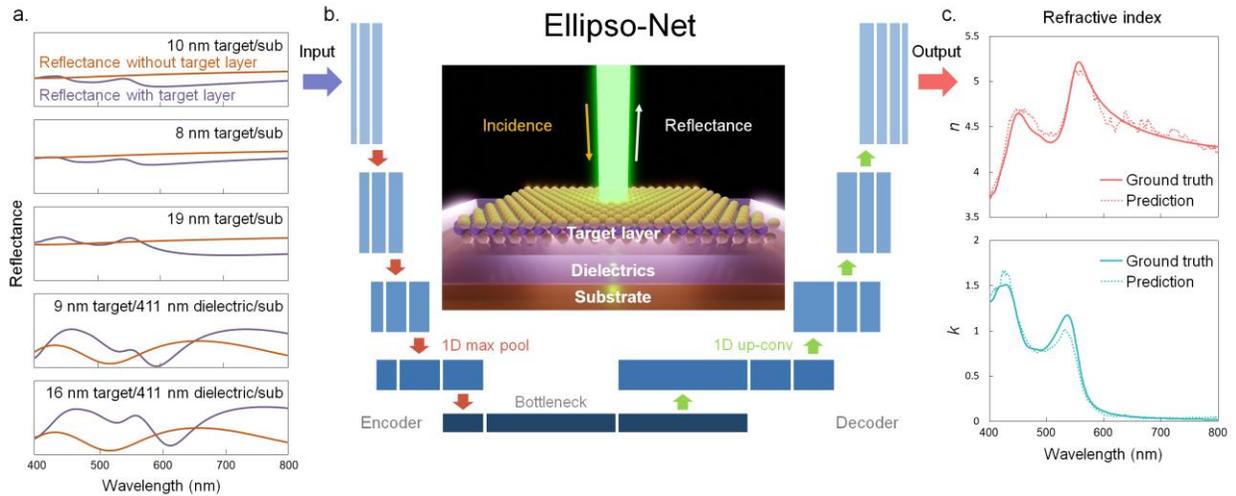

**Figure 1. Overall workflow of refractive index prediction.** (a) Input of 5 pairs of reflectance spectra measured with and without the target layers. (b) Schematic illustration of the encoder-decoder convolutional neural network, EllipsoNet, with multiple convolutional layers (blue rectangles), four down-sampling layers (red arrows), and four up-sampling layers (green arrows). The inset shows a schematic of the multilayer stack structure used in this study. (c) Prediction of the real part, *n*, and the imaginary part, *k*, of refractive indices in dash curves and their corresponding ground truths in solid curves.

## RESULTS AND DISCUSSION

The test structure is a multilayer stack consisting of a thin film of interest (target layer), a relatively transparent dielectric layer, and a thick substrate layer (illustrated in the inset of **Figure 1b**). The thicknesses and the complex refractive indices of the dielectric and the substrate layers are generated randomly and "unknown" to the training model. The input data of the learning problem are 5 pairs of reflectance spectra measured or simulated from 5 different multilayer stacks with the same material (the same refractive indices) in the target layer, but different materials in both the dielectric layer and the substrate layer. In addition, the thicknesses of the target layer and dielectric layer can be different (see **Table S1** for details). Each pair of reflectance spectra contains two reflectance spectra for the same multilayer stack structure, but one with the target layer and the other without the target layer. **Figure 1a** shows one example of the input data. The outputs of the learning problem are complex refractive indices of the target material as a function of the incident light wavelength (**Figure 1c**). The spectral range selected in this study is the visible range (400 nm – 800 nm), but the method can be readily applied to any spectral range of interest. The model is based on an encoder-decoder convolutional neural network named *EllipsoNet*, as shown in **Figure 1b**. The downsampling (encoder) and upsampling (decoder) convolutional layers are symmetric. This structure is designed to better extract spectroscopic features in the input data and produce same-dimensional refractive index spectra as the outputs. Details about EllipsoNet can be found in the Experimental Section, **Figure S3,** and **Table S2**.

Deep learning models typically require large training datasets. However, producing training data in large quantities through experiments is extremely time-consuming. To address this issue, the datasets (including training, validation, and testing) used in this study were numerically simulated (**Figure S4**). The refractive indices of the target layer were extracted from the C2DB database,[15–17] which are theoretically simulated by the DFT with GW approximation and the Bethe-Salpeter equations (BSE).[18] 400 different materials from C2DB were used and further augmented through a transformation satisfying the KK relations. The refractive indices of the dielectric and substrate layers were generated using the Sellmeier formula.[19] The multilayer stack structures (materials and thicknesses) were randomly generated, and the optical reflectance spectra were computed using the transfer matrix method.[20] With the help of the numerical database and additional data augmentation approaches (see the Experimental Section for additional information), reflectance spectra of 450,000 different multilayer stacks were generated and grouped into 90,000 input-output pairs: 10 reflectance spectra of 5 multilayer stack structures as each input data, and the refractive indices of materials in the target layer as the ground truth of the corresponding predictions ("labels"). 95% of the input-output pairs were used as the training dataset, and 5% were used as the validation dataset. Additionaly, we generated an independent testing dataset that contains 10,000 multilayer stack structures using completely different target layer materials ("unseen" by the learning model). To train the model, we developed a loss function that combines the mean square error (MSE) and 1 minus Pearson's correlation coefficient (1−PCC). 1−PCC ensures the fast convergence of the positions, shapes of the characteristic peaks, and other relative spectral features. MSE minimizes the discrepancy of the absolute values at each wavelength. An Adam optimizer with an exponential learning rate decay scheduler was adapted during the training.[21] Details about the convolutional neural network architecture, the loss function, and the training procedure are described in the Experimental Section.

In order to measure the prediction accuracy of the fully trained EllipsoNet, we introduced two metrics: the root mean square percentage score (RMSPS) and Pearson's correlation coefficient (PCC). RMSPS is sensitive to the absolute difference between the ground truth and the prediction, and is in the range of $-\infty$ to 1, where 1 corresponds to perfect prediction and a lower value means less accurate prediction. PCC measures the similarities of the shapes between the ground truth and the prediction, and is in the range of $-1$ to 1, where 1 corresponds to perfect prediction. Details about these two metrics are discussed in the Experimental Section. **Figure 2a** shows examples of the predicted complex refractive indices. These examples are selected randomly from each dataset. The red curves are the real part *n*, and the green curves are the imaginary part *k*. The predicted curves (dashed) match very well with the ground truth (solid), both in terms of spectral features and absolute values. **Figure 2b** summarizes the RMSPS and the PCC of the training, the validation, and the testing datasets. Our trained EllipsoNet achieves a median RMSPS of 0.86 and a median PCC of 0.93 on the training set, and a median RMSPS of 0.83 and a median PCC of 0.91 on the validation set. Both the RMSPS and the PCC are similar on training and validation sets, indicating a good convergence of training without overfitting. Both training and validation sets reach the highest RMSPS of 0.97 and the highest PCC of 0.99.

With the testing set generated independently, a median RMSPS of 0.81 and a median PCC of 0.88 are reached, as shown in the last panel in **Figure 2b**. The highest RMSPS and PCC for the testing set are 0.95 and 0.99, respectively. These scores are only slightly lower than those on training and validation sets, confirming the fidelity and robustness of the predictions made by EllipsoNet.

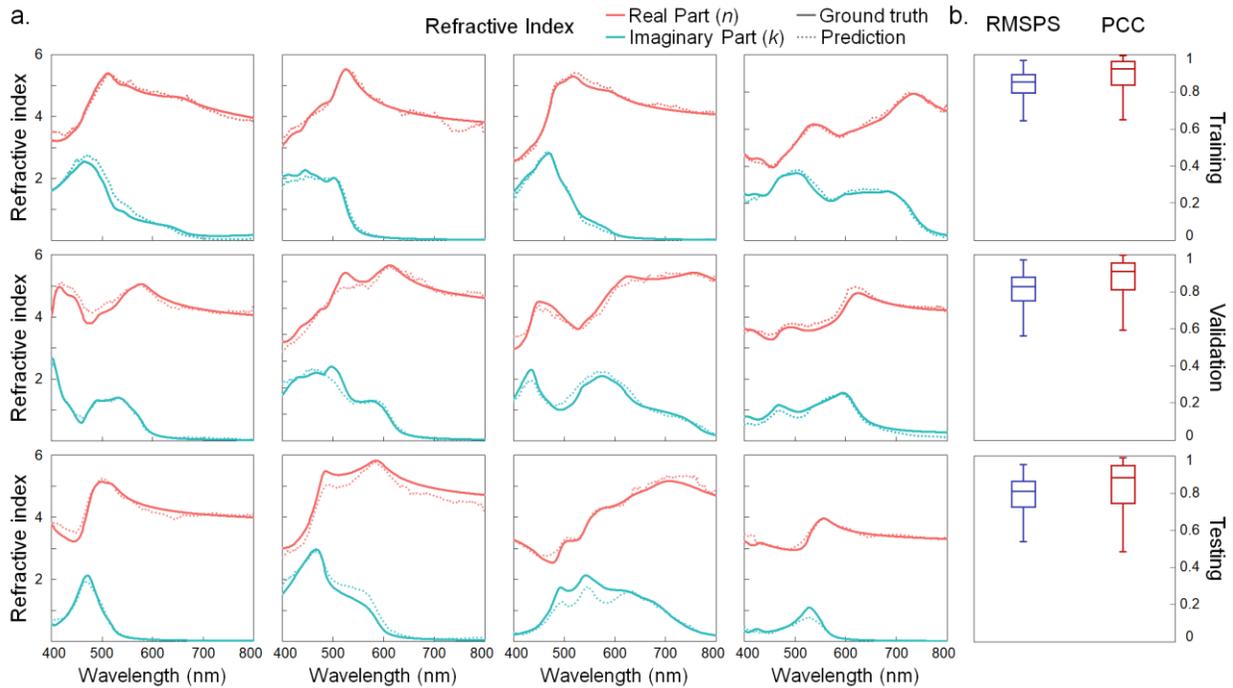

**Figure 2. Predictions of complex refractive indices made by EllipsoNet on the training, the validation, and the testing sets.** (a) Examples of the predicted $n$ and $k$ in dash curves and their corresponding ground truths in solid curves tested on the training (top row), the validation (middle row), and the testing sets (bottom row). (b) Box plots showing 0, 25, 50, 75, and $100^{th}$ percentiles of the RMSPS (blue) and the PCC (red) for the training, the validation, and the testing sets.

We further demonstrate that EllipsoNet could make predictions with reasonable accuracies from experimentally measured data, even though the model is trained on numerically simulated data. 2D material thin flakes, including $MoS_2$, $MoSe_2$, $WS_2$, and $WSe_2$ with different thicknesses, were exfoliated onto two different substrates: bare silicon and 300 nm $SiO_2$/silicon substrates. Although our model can be applied to all kinds of substrate structures, we chose these two substrates because they are the most commonly used substrates for 2D material studies. All of the four 2D materials, even the simulated ones, are not seen by EllipsoNet during training. In our experiment, $MoS_2$ flakes with thicknesses of 5 nm, 10 nm, and 26 nm on bare silicon substrates and $MoS_2$ flakes with thicknesses of 3 nm and 9 nm on 300 nm $SiO_2$/Si substrates were prepared. The thicknesses of the target flake were confirmed by atomic force microscopy. The corresponding reflectance spectra and optical microscopic images are shown in **Figure 3a** and **b**. We also measured the reflectance spectra of empty substrates without $MoS_2$ flakes (without the target layer), as shown in the red curves in **Figure 3a**. Similarly, we prepared testing structures and measured reflectance spectra for additional $MoS_2$ samples and for $MoSe_2$, $WS_2$, and $WSe_2$ (**Figure S5 – S8**). The reflectance spectra of these samples were taken by a benchtop optical microscope coupled with a broadband light source and a spectrometer, and were fed into the fully trained EllipsoNet for the prediction of the complex refractive indices. EllipsoNet predicts both $n$ and $k$ as shown in dashed curves in **Figure 3c** and **Figure 3d.** For all the four 2D materials, the predicted $n$ and $k$ (dashed curves) match reasonably well with the ground truth (solid curves) in terms of characteristic peaks, trends, and absolute index values. The prediction accuracies for the experimental results are slightly lower than those for the numerically generated data, possibly because there are discrepancies in the data distributions (trends and absolute index ranges) between the experimental dataset and the numerically generated dataset. To alleviate this issue, we have developed an approach to augment and transform the refractive indices from the simulated database to match the distributions with the experimental dataset better (detailed discussion can be found in the Experimental Section). These results indicate that our EllipsoNet trained with numerically generated data can achieve high performance in predicting real experimental results.

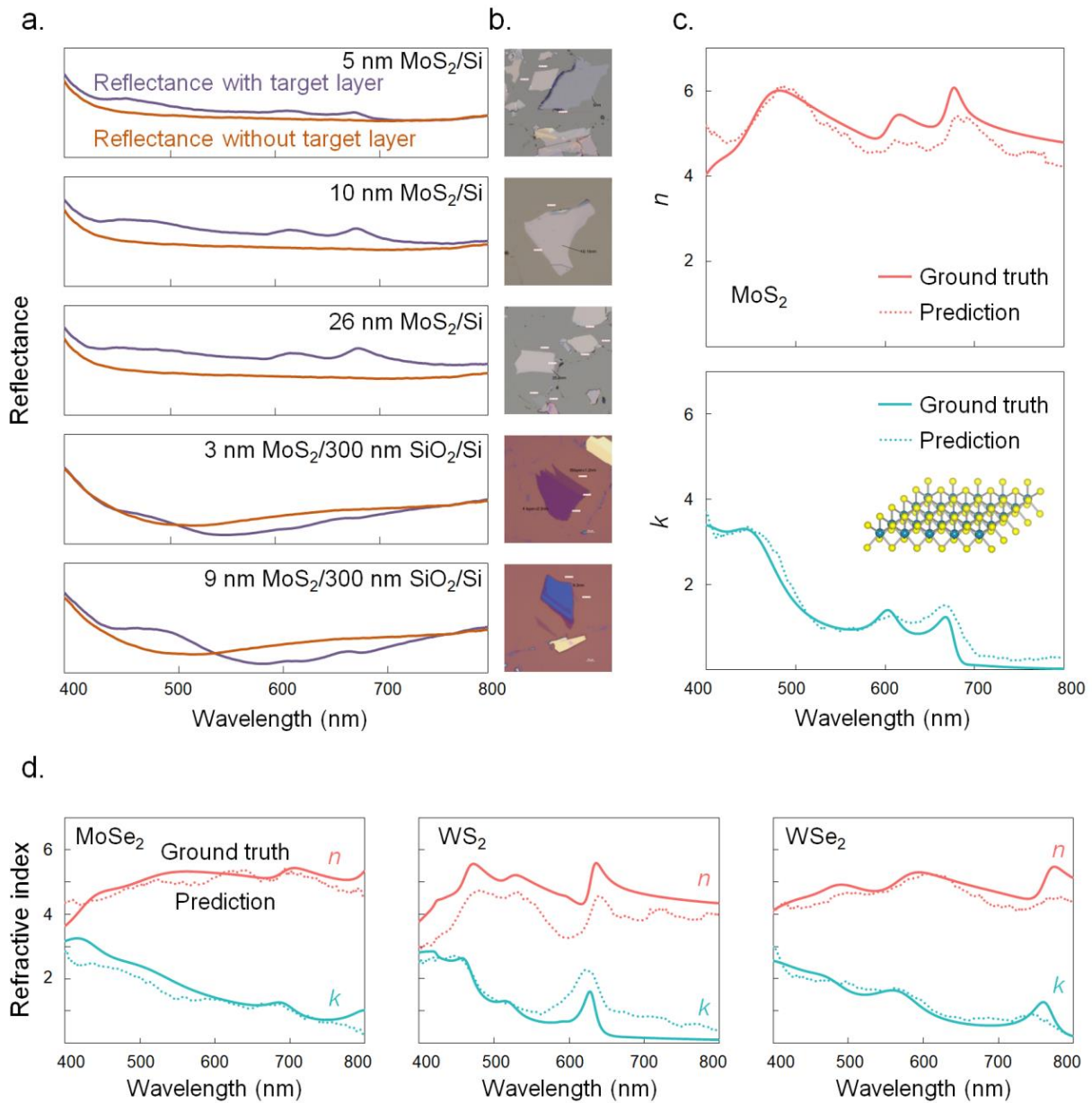

**Figure 3. Predictions of complex refractive indices on experimentally measured reflectance spectra for exfoliated 2D materials.** (a) 5 pairs of reflectance spectra of different multilayer stack structures with and without the exfoliated $MoS_2$ as the target layer. (b) The corresponding optical microscopic images of $MoS_2$ with different thicknesses, and on bare silicon (first three rows) and

on 300 nm $SiO_2$/Si substrates (bottom two rows). (c) Predictions of $n$ and $k$ in dash curves and their corresponding ground truths in solid curves for the $MoS_2$ samples. (d) Predictions of $n$ and $k$ in dash curves and their corresponding ground truths in solid curves for $MoSe_2$, $WS_2$, and $WSe_2$ samples.

A unique advantage of EllipsoNet is that it can extract complex refractive indices of a thin film placed on top of a nontrivial substrate with unknown information. In the previous section, we trained EllipsoNet with simple multilayer stack structures (one target material layer, none or one dielectric layer, and one substrate layer), denoted as S-structure. However, 2D materials in photonic and electronic devices are sometimes covered by passivation or native oxide layers. Also, 2D materials are often placed on specifically designed, more complicated multilayer dielectric structures. To consider more practical applications, we trained another EllipsoNet with the target material layers covered by top dielectrics and placed them on more complex multilayer stack structures. We call this new model complex-EllipsoNet, or C-EllipsoNet in short. The detailed stack structures for C-EllipsoNet, called C-structure, is shown in Table S1. We generated a new dataset numerically with 450,000 multilayer stacks with 0 or 1 layer of dielectrics above the target material layer (top dielectric stack, TDS) and 0 to 3 layers of dielectrics under the target material layer (bottom dielectric stack, BDS). The complex multilayer stack structures with TDS, target layer, BDS, and substrate are illustrated in **Figure 4a,** and details are described in the Experimental

Section. We obtained C-EllipsoNet by using the same training setup as previously described for EllipsoNet in the Experimental Section. C-EllipsoNet achieves high performance in predicting refractive indices of thin films from optical reflectance spectra when placed on complex, unknown multilayer stack structures. Predictions are made by both EllipsoNet and C-EllipsoNet on a series of newly generated testing sets with the materials of the target layers "unseen" by the models. In these new testing sets, the level of complexities of the multilayer stack structures are varied, and the prediction results are summarized in **Figure 4**.

First, the performance of EllipsoNet and C-EllipsoNet for the predictions of testing datasets with different levels of complexities are compared, as shown in **Figure 4b**. Again, EllipsoNet is trained by an S-structure training set, and C-EllipsoNet is trained by a C-structure training set. Here we tested EllipsoNet on the S-structure testing set, denoted as SS (simple on simple) in **Figure 4b**; and test C-EllipsoNet on both the S-structure and the C-structure testing set, denoted as CS (complex on simple) and CC (complex on complex), respectively. We noticed that the median of RMSPS is slightly improved from 0.81 to 0.84, and the median of PCC is slightly improved from 0.88 to 0.89 when the testing scenario is changed from SS to CC, whereas the variations of both RMSPS and PCC are increased. These results indicate improved overall scores of the prediction accuracies, but at the same time, the accuracies also have broader distributions. On the other hand, both RMSPS and PCC are dropped in the CS testing scenario. In summary, the performances of C-EllispoNet (CC), EllipsoNet (SS), and C-EllispoNet with simpler structures (CS) can be ordered as CC > SS > CS. To better visualize the trend, we defined the wavelength-dependent MSEs between the test results of certain testing scenarios (T = SS, CC, or CS) and the ground truth (G) values of *n* and *k*, denoted as MSE(T, G). As shown in **Figure 4d** and **Figure 4e**, in predicting both *n* and *k*, MSE(CC, G) is the smallest across all wavelengths, indicating a better performance of CC. **Figure 4g, h, j, k** are two examples of the predicted *n* and *k* as a function of wavelength by SS, CC, and CS compared with the ground truth. The differences in performance with different training structures show that our framework can work even better in predicting refractive indices of thin films in more complex stack structures than in simple ones.

Further, we investigated the impact of the level of complexities of the multilayer dielectric structures on the prediction accuracies made by C-EllipsoNet, and the key metrics are summarized in **Figure 4c**. We systematically varied the level of complexities in the testing dataset, including the number of layers in TDS, the number of layers in BDS, the number of different types of materials in BDS in each group (5 pairs of the reflectance spectra) of the input data, and the number of different types of materials in the substrate layer in each group of the input data. The key observations are: (1) RMSPS and PCC are less affected by the number of layers in TDS (TDS # 1, 0 in **Figure 4c**); (2) the performance of C-EllipsoNet degrades only slightly (the median RMSPS and the median PCC decrease from 0.83 to 0.8 and from 0.9 to 0.87, respectively) when the number of layers in BDS decreases from 3 to 1 (BDS # 3, 2, 1 in **Figure 4c**); (3) a significant drop of both the median RMSPS and the median PCC is observed when the number of material types in the bulk substrate in each group of the input data is decreased from 5 to 1 (BDS # 1 to 1' in **Figure 4c**): the median RMSPS and the median PCC drop from 0.8 to 0.76 and from 0.87 to 0.84, respectively; (4) when the number of different types of materials in BDS in each group of the input data decreases from 5 to 1 (BDS # 1' to BDS variation 1 in **Figure 4c**), the performance of C-EllipsoNet also decreases slightly, with the median RMSPS changing from 0.76 to 0.75, and the median PCC changing from 0.84 to 0.81; and (5) a much more severe performance degradation takes place when no BDS is present in the structure (the median RMSPS and the median PCC decrease from 0.75 to 0.69 and from 0.81 to 0.8, respectively; BDS variation 0 in **Figure 4c**). In summary, the prediction framework shows higher prediction accuracies for more complex multilayer stack structures, probably because the model is given access to a much wider hyperspace of variables in this physical problem in the training stage. This aspect is fundamentally more advantageous than traditional methods such as optical ellipsometry with deterministic model fittings, which can only extract the refractive indices of a thin film on a known simple substrate structure. We thus envision that our approach can be further developed to predict optical properties of materials in-operando when they are integrated into complex photonic structures and electronic devices.

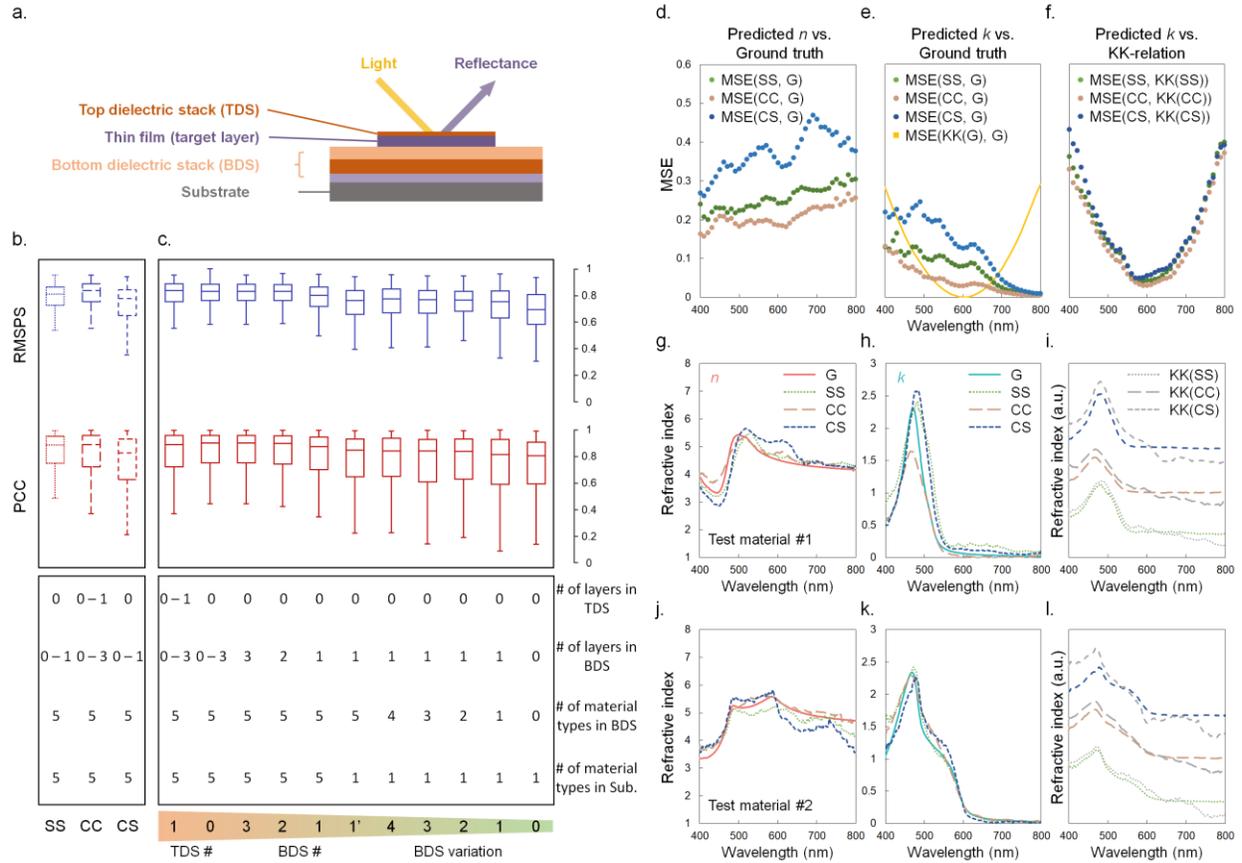

**Figure 4. Predictions of complex refractive indices on complex multilayer stack structures, and the learning of the KK-relations.** (a) Schematic of the C-structure with TDS, target layer, BDS, and substrate. (b) Box plots showing 0, 25, 50, 75, and 100th percentiles of RMSPS and PCC performances using different models and testing sets. The associated table summarizes the key parameters in each testing scenario. SS (simple on simple) is EllipsoNet tested on S-structure; CC (complex on complex) is C-EllipsoNet tested on C-structure; and CS (complex on simple) is C-EllipsoNet tested on S-structure. (c) Box plots showing 0, 25, 50, 75, and 100th percentiles of RMSPS and PCC performances using C-EllipsoNet and tested on testing sets with structures of different levels of complexities. The associated table summarizes the key parameters of different

test structures. TDS # is the number of the top dielectric layers. BDS # is the number of the bottom dielectric layers. BDS # 1' is test strcutres with 1 dielectric layer in BDS, 5 types of material in BDS and 1 type of material in substrate. BDS variation is the number of different types of materials in BDS in each group of the input data. The testing structures are ordered by decreasing complexities from left to right. (d) Wavelength-dependent MSEs of ground truth $n$ and model-predicted $n$ for SS, CC, and CS. (e) Wavelength-dependent MSEs of ground truth $k$ and model-predicted $k$ for SS, CC, and CS. (f) Wavelength-dependent MSEs between KK-relation-calculated $k$ and model-predicted $k$ by SS, CC, and CS. (g) Predicted $n$ as a function of wavelength by SS, CC, and CS compared with the ground truth of test material #1. (h) Predicted $k$ as a function of wavelength by SS, CC, and CS compared with the ground truth of test material #1. (i) KK-relation-calculated $k$ compared with model-predicted $k$ of test material #1. The other curves are the same as SS, CC, and CS in (h). (j) Predicted $n$ as a function of wavelength by SS, CC, and CS compared with the ground truth of test material #2. (k) Predicted $k$ as a function of wavelength by SS, CC, and CS compared with the ground truth of test material #2. (l) KK-relation-calculated $k$ compared with model-predicted $k$ of test material #2. The other curves are the same SS, CC, and CS in (k).

Finally, we demonstrated that our EllipsoNet was capable of learning not only the specific data but also the inherent physics buried in the data. The real and imaginary parts of the complex refractive indices of materials are governed by the KK relation, which results from the causality

Page **18** of **32**

of electromagnetism. Both EllipsoNet and C-EllipsoNet spontaneously learn the KK relations between *n* and *k* from the training dataset and the *n* and *k* values predicted by our models automatically satisfy the KK relations. We used the KK relations to compute *k* values from model-predicted *n* values. Details on the calculations of refractive indices using the KK relations are described in the Experimental Section. **Figure 4f** shows that the MSEs between the model-predicted *k* values and the KK-relation-calculated *k* values are reasonably small in the entire spectral range, denoted as MSE(T, KK(T)), where T stands for SS, CS, or CC. All three MSEs are lower at the center wavelength (600 nm) and higher at the two boundaries of the given spectral range. In **Figure 4e**, we plotted MSE(KK(G), G) for *k* values, where KK(G) represents the *k* values computed from the ground truth *n* values through the KK relations, as the yellow solid curve. MSE(KK(G), G) has the same trend. The pattern is caused by the integrals in the KK relations, which require the *n* values for the entire spectral range even for computing any single wavelength *k* value. Therefore, extrapolation outside the given spectral range needs to be made, and the computed *k* values are inevitably inaccurate near the boundaries of the given spectral range. **Figure 4i** and **Figure 4l** are two material examples (the y-axes are shifted for each testing scenario for visual clarity). The predicted *k* values and the KK-relation-calculated *k* values match reasonably well.

Moreover, EllipsoNet and C-EllipsoNet can make predictions about *n* and *k* with better accuracies than the KK-relation-computed values around the boundaries of the given spectral range. In **Figure 4e**, compared to MSE(KK(G), G), all the model-predicted *k* values (SS, CS, CC) have smaller MSEs around both boundaries of the given spectral range. Both models can better learn the relationships between *n* and *k* from optical reflectance spectra, outperforming the KK-relation-calculated values, especially when the provided spectral range is limited.

**CONCLUSION**

In summary, we developed a computational ellipsometry approach based on a deep learning model that extracts complex refractive indices of thin films on complicated, unknown substrates with simplified optical setups. Specifically, we designed an encoder-decoder convolutional neural network named EllipsoNet that takes the reflectance spectra measured on a desktop optical microscope as the inputs and predicts the corresponding refractive indices of the target thin film materials. The model is trained on numerically generated data sets but can be used for experimentally measured data. We further demonstrated that the approach could extract refractive indices of materials in relatively complex multilayer material stacks, and the model can spontaneously learn the underlying physical principles, the Kramers-Kronig relations. This method can readily be used for in-operando optical characterizations of functional materials in complex photonic devices and can be extended to other spectroscopic characterization applications. Our work is a salient demonstration showing that machine learning can simplify material characterization: using simple characterization and judiciously developed machine learning models, we can obtain material properties that traditionally require complicated instrumentation.

# EXPERIMENTAL SECTION

## Database extraction and preparation

We extracted the optical properties of 2D materials from a computational 2D material database (C2DB).[16,17] Firstly, we developed a web crawler to download optical polarizabilities of over 400 DFT simulated materials. We used the polarizability along x in energy level range 0 to 10 eV, which is set in the database for further calculations. To convert the polarizabilities into refractive indices, we used the equation as follows:

$$\eta(\omega) = \sqrt{\chi(\omega) + 1} \qquad (1)$$

Here, the polarizability $\chi(\omega)$ and refractive index $\eta(\omega)$ are complex numbers at wavelength $\omega$.

## Reflectance calculation with data augmentation

We randomly generated stacks of materials. With randomly generated structures of thin films placed on substrates, we used the transfer matrix method implemented in MATLAB to calculate the reflectance spectra from the refractive indices of the materials.[20] Specifically, in the multilayer stack structures, dielectric layers and the target thin film layer are treated differently. For the target thin film layer, we augmented the refractive indices calculated from C2DB using the following transformations:[22]

$$\epsilon_1'(\omega) = \epsilon_1(\omega) \times c_1 + c_2 \tag{2}$$
$$\epsilon_2'(\omega) = \epsilon_2(\omega) \times c_1 \tag{3}$$

Here, $\epsilon_1$ and $\epsilon_2$ are real and imaginary parts of complex dielectric function in the range of 400 nm to 800 nm. The complex dielectric functions are related to the refractive index by $\epsilon_1(\omega) = n(\omega)^2 - k(\omega)^2$ and $\epsilon_2(\omega) = 2n(\omega) * k(\omega)$. Using different $c_1$ and $c_2$, we were able to generate more materials. Such transformations satisfy the KK relations, and detailed proof can be found in Supporting Information. The ranges of $c_1$ and $c_2$ were chosen to make the DFT simulated materials better reflect the distributions of complex refractive indices of real 2D materials. The thicknesses of the target layers are varied in the range of 0.3 – 20 nm (**Table S1**).

To generate the transparent materials ($k = 0$) in the dielectric layers, we used the Sellmeier formula as follows:[19]

$$n(\lambda) = \sqrt{1 + \frac{c_1 \times \lambda^2}{\lambda^2 - c_2^2} + \frac{c_3 \times \lambda^2}{\lambda^2 - c_4^2} + \frac{c_5 \times \lambda^2}{\lambda^2 - c_6^2}} \tag{4}$$
$$k(\lambda) = 0 \tag{5}$$

Here, the $c_1, c_2, c_3, c_4, c_5$ and $c_6$ are variables for different materials, and $\lambda$ is wavelength in micrometers. For example, the constants for fused silica are $c_1 = 0.696$, $c_2 = 0.068$, $c_3 = 0.408$, $c_4 = 0.116$, $c_5 = 0.897$ and $c_6 = 9.896$. We set $c_1 = 0.5 - 2$, $c_2 = 0.004 - 0.0095$, $c_3 = 0 - 0.7$, $c_4 = 0.08 - 0.16$, $c_5 = 0.8 - 0.68$ and $c_6 = 9 - 16$ with uniform distribution, according to various dielectric materials in an experimental refractive index database.[23] The thicknesses of dielectric materials are in the range of 10 – 500 nm.

As shown in **Figure 4a**, the dielectric materials placed on thin film are named TDS, and dielectric materials placed under thin film are named BDS. There is always one layer of thick dielectric substrate with infinite thickness named substrate. Such a thick layer is necessary for realistic experimental samples but needs to be treated separately in the transfer matrix method calculation because the constraint for phase coherence needs to be removed.

**Neural network model and training procedure**

The input data dimension of EllipsoNet is $10 \times 400$. There are 10 channels of 1D vectors with 400 features in each channel, representing 5 pairs of reflectance spectra, in the wavelength range of 400 nm to 800 nm, of multilayer stack structures with and without the target layer. The output data dimension of the network is $2 \times 400$. These 2 channels are 1D vectors with the same length as the input vectors, representing *n* and *k*, respectively, of the target layer thin film material in the same wavelength range. As shown in **Figure 1**, **Figure S3,** and **Table S2**, the input first goes through a 1D convolutional layer with 10 input channels, 64 output channels, the kernel size of 3, the stride of 1, and the padding of 1 to maintain the same feature dimensions. Then, there is one 1D batch normalization layer and one rectified linear unit (ReLU) activation layer.[24] There are two more 1D convolutional layers with the same input and output channel numbers followed by 1D batch normalization and ReLU. These three 1D convolutional layers form a three-convolutional-layers group with the input dimension c × f and the output dimension 2c × f, where c is the number of channels and f is the number of features. Then, one 1D max pooling layer with the kernel size of 2 and the stride of 2 is applied to reduce the feature dimensions by half. Including the first convolutional layer group followed by max pooling, there are four of such three-convolutional-layer groups followed by a max pooling layer with doubled channels and halved features each time. After the encoder section, there are 512 channels with 50 features in each channel. Then, there is a bottleneck layer with the output of 1024 channels with 25 features in each channel, followed by one dropout layer with a dropout rate of 0.5. In the decoder section, there are four up-sample layers followed by three-convolutional-layers groups with input $2c \times f$ and output $c \times f$. Similar to the encoder section, each three-convolutional-layers group consists of 3 1D convolutional layers: the first 1D convolutional layer is with the number of output channels half of the number of the input channels, the kernel size of 3, the stride of 1, and the padding of 1 followed by one 1D batch normalization layer and ReLU; and $2^{nd}$ and 3rd 1D convolutional layers are with the input and output channels the same size, followed by 1D batch normalization and ReLU. After the decoder section, one convolutional layer with 2 output channels, the kernel size of 1, the stride of 1, and the padding of 0 is applied as the output layer.

For the training of EllipsoNet and C-EllipsoNet, we generated reflectance spectra pairs of 450,000 multilayer stack structures with and without the target layer. We combined every 5 pairs of reflectance spectra as an input group. There are 90,000 input data groups in total. We randomly split the dataset into training and validation sets and generated a testing set independently with thin films unseen in the other datasets. We also added gaussian noise to the input reflectance spectra in training to prevent overfitting. Using the training set, we trained the network with the Adam optimizer and the exponential learning rate decay scheduler till it converged (**Figure S9**, **S10**).[21]

We chose the MSE loss combined with 1−PCC implemented by the cosine similarity between the mean-centered prediction and ground truth as the loss function, described as following: [25]

$$loss(Y, \hat{Y}) = \frac{1}{n} \sum (\frac{1}{2}(Y_{ni} - \hat{Y}_{ni})^2 + \frac{1}{2}(Y_{ki} - \hat{Y}_{ki})^2)$$
$$+ 10 \times (1 - \frac{\sum(Y_{ni} - \mu_n)(\hat{Y}_{ni} - \hat{\mu}_n)}{\sqrt{\sum(Y_{ni} - \mu_n)^2}\sqrt{\sum(\hat{Y}_{ni} - \hat{\mu}_n)^2}} + 1 - \frac{\sum(Y_{ki} - \mu_k)(\hat{Y}_{ki} - \hat{\mu}_k)}{\sqrt{\sum(Y_{ki} - \mu_k)^2}\sqrt{\sum(\hat{Y}_{ki} - \hat{\mu}_k)^2}})$$
(6)

Here, $Y_n$ is the ground truth $n$ and $\hat{Y}_n$ is the predicted n. $Y_k$ is the ground truth $k$ and $\hat{Y}_k$ is the predicted k. $\mu_n$ is the mean of the ground truth $n$ and $\hat{\mu}_n$ is the mean of the predicted n. $\mu_k$ is the mean of the ground truth $k$ and $\hat{\mu}_k$ is the mean of the predicted k. Specifically, the MSE between the ground truth and the predicted $n$ and $k$ (the first term) minimizes the difference in the absolute refractive index values. The 1−PCC between the ground truth and the predicted $n$ and $k$ (the second term) minimizes the relative characteristic trends, peaks, and features.

Different setups of model complexity, loss function selection, and noise configuration have been tested in the training of S-structure while all the other hyperparameters, including epoch, learning rate decay, mini-batch size, *etc.*, are fixed. First, models with different structures are trained. When the bottleneck layer (the deepest layer between encoder and decoder) is added, the median PCC for the training set is significantly improved, from 0.84 to 0.87. However, the median PCC for the validation set drops from 0.82 to 0.78 and has broader distributions, indicating overfitting. To prevent overfitting, a dropout layer with a dropout rate of 0.5 is added after the bottleneck layer.[26] Both the median PCC for training and validation sets are improved (to 0.89 and 0.88, respectively), and the performance is more robust (narrower distributions). Then, a gaussian error linear unit (GELU), with a smooth region around zero and gradients in all regions, is used to replace ReLU.[27] Both the median PCC for training and validation sets slightly drop. Finally, skip connections between the convolutional layers in the same depth of encoder and decoder are added to enable reusability of the features similar to those used in the U-net.[28] Again, both the median PCC for training and validation sets slightly drop. Therefore, in the following model training, an encoder-decoder convolutional neural network with bottleneck layer, dropout layer, and ReLU activation is used. Second, different loss functions are used in training. The MSE measures the average of squares of errors and is used to minimize the discrepancy of absolute values at each wavelength. When using only MSE as the loss, the model achieves a median PCC of 0.89 and 0.88 for training and validation sets, respectively. Then, the 1−PCC is added to ensure the fast convergence of the positions and shapes of the characteristic peaks as well as other relative spectral features. Both the median PCC for training and validation sets are improved, to 0.91 and 0.9. After a scaling factor is applied to the 1−PCC to balance the different scales between MSE loss and 1−PCC loss, the median PCC for training and validation sets are boosted to 0.93 and 0.91. In all the setups, gaussian noises are added to the inputs in training. When we remove the artificial noise in training, the median PCC of the training set increases from 0.89 to 0.91, while the median PCC of the validation set stays at 0.88, indicating severer overfitting. A detailed performance summary of training using different model structures, loss functions, and noise is in **Figure S11**.

**Performance evaluation criteria**

To check the prediction performance of our model, we used two metrics, RMSPS and PCC. Both parameters represent perfect predictions when reaching 1. RMSPS is 1 minus the mean squared error normalized by the mean of the ground truth refractive indices. If the MSE is 0, RMSPS will be 1. If the MSE exceeds the mean of the ground truth, RMSPS will become negative. PCC is commonly used in measuring the correlation between two variables. The mathematical formulations of these two performance metrics are:

$$RMSPS = 1 - \frac{\sqrt{\Sigma(Y_{ni} - \hat{Y}_{ni})^2 + (Y_{ki} - \hat{Y}_{ki})^2}}{\Sigma(Y_{ni} + Y_{ki})} \tag{7}$$

$$PCC = \frac{\Sigma(Y_{ni} - \mu_n)(\hat{Y}_{ni} - \hat{\mu}_n)}{2\sqrt{\Sigma(Y_{ni} - \mu_n)^2}\sqrt{\Sigma(\hat{Y}_{ni} - \hat{\mu}_n)^2}} + \frac{\Sigma(Y_{ki} - \mu_k)(\hat{Y}_{ki} - \hat{\mu}_k)}{2\sqrt{\Sigma(Y_{ki} - \mu_k)^2}\sqrt{\Sigma(\hat{Y}_{ki} - \hat{\mu}_k)^2}} \tag{8}$$

**Experimental sample preparation and spectroscopy measurements**

2D material thin flakes were mechanically exfoliated by Scotch tape from bulk single crystals of $MoS_2$, $MoSe_2$, $WS_2$, and $WSe_2$ (HQ Graphene) onto bare Si and $SiO_2$/Si ($SiO_2$ thickness = 300nm) substrates, respectively. Flakes were identified and located by optical microscope, and the thickness of each flake was determined by atomic force microscope (AFM) (Bruker, Dimension Icon).

The micro-reflection spectroscopy was measured on a home-built spectroscopy system using Laser-Driven Light Sources as the broadband white light source. The white light was incident from a 100× objective (NA = 0.9) with a beam size of 8 μm. The spectrometer was coupled with a 600/mm grating and a Horiba Syncerity CCD detector to measure the reflection in the visible wavelength.

The ellipsometry measurement was conducted on a Woollam M-2000F focused beam spectroscopic ellipsometer. The measuring angle was fixed at 65°, and the focused spot size was around 25 × 60 μm. The measured flakes were located using a digital CCD camera. CompleteEASE software was used to perform the data analysis to obtain the dielectric constant from the measured data. The fitting was done using the Tauc-Lorentz model.

**Kramers-Kronig relation calculation**

We implemented the calculation of the KK relations with a numerical implementation of integration and differentiation using the equations as follows:

$$n(\omega) = 1 + P \int_{-\infty}^{+\infty} \frac{d\omega'}{\pi} \frac{k(\omega')}{\omega' - \omega} \tag{9}$$

$$k(\omega) = -P \int_{-\infty}^{+\infty} \frac{d\omega'}{\pi} \frac{n(\omega') - 1}{\omega' - \omega} \tag{10}$$

To calculate the $k$ for each wavelength $\omega$, we used backward differentiation for $d\omega'$ and nearest extrapolation for wavenumbers below and beyond the known wavenumbers to achieve $\pm\infty$ in the integral. Then the KK calculated $k$ is calibrated by the mean of the ground truth $k$ by subtracting the difference between the averages.

The numerical integration in the range 400 nm to 800 nm with extrapolation outside the range is as follows, where $f(k) = \frac{2c\pi}{k}$ converges the wavelength k to angular frequency $\omega$:

$$n(\omega) = 1 + \frac{1}{\pi} \sum_{i'=400}^{800} \frac{k(f(i'))}{f(i') - \omega} \times (f(i') - f(i'-1)) + \sum_{i'=2}^{399} \frac{k(f(400))}{f(400) - \omega} \times (f(i') - f(i'-1)) + \sum_{i'=801}^{1200} \frac{k(f(801))}{f(801) - \omega} \times (f(i') - f(i'-1)) \quad (11)$$

$$k(\omega) = -\frac{1}{\pi} \sum_{i'=400}^{800} \frac{n(f(i'))-1}{f(i') - \omega} \times (f(i') - f(i'-1)) - \sum_{i'=2}^{399} \frac{n(400)-1}{f(400) - \omega} \times (f(i') - f(i'-1)) - \sum_{i'=801}^{1200} \frac{n(801)-1}{f(801) - \omega} \times (f(i') - f(i'-1)) \quad (12)$$

**ASSOCIATED CONTENT**

**Supporting Information**

Supporting Information is available from the Wiley Online Library or from the author.

**AUTHOR INFORMATION**

**Author Contributions**

Z. W., Y.C.L., and S.H. conceived this work and designed the algorithm. Z.W. implemented the algorithms and performed the model training. Y.C.L., Z.W., and S.H. designed and performed data analysis. K.Z. measured reflectance and ellipsometry spectra. W.W. prepared thin film samples. Z. W., Y.C.L., and S.H. wrote the manuscript with the inputs from all the authors.

**Funding Sources**

Z.W. and S.H. acknowledge the support from the National Science Foundation under grant number ECCS-1943895.